\documentclass[journal]{IEEEtran}

\ifCLASSINFOpdf
\else
   \usepackage[dvips]{graphicx}
\fi
\usepackage{url}

\usepackage{graphicx}
\usepackage{multirow}
\usepackage{amssymb}
\usepackage{xcolor}
\usepackage{cite}
\usepackage{amsthm}
\usepackage{amsmath}
\usepackage{booktabs}

\usepackage{amssymb}
\usepackage{pifont}
\newcommand{\cmark}{\ding{51}}%
\newcommand{\xmark}{\ding{55}}%

\begin{document}

\title{Selective HuBERT: Self-Supervised Pre-Training for Target Speaker in  Clean and Mixture Speech}


\author{Jingru Lin, Meng Ge$^{*}$, Wupeng Wang, 
Haizhou Li, \IEEEmembership{Fellow, IEEE}, Mengling Feng, \IEEEmembership{Senior Member, IEEE}
\thanks{Jingru Lin and Wupeng Wang, Department of Electrical and Computer and Engineering, National University of Singapore, Singapore 119077 (e-mail: jingrulin@u.nus.edu; e0125301@u.nus.edu). 

Meng Ge and Mengling Feng, Saw Swee Hock School of Public Health, National University of Singapore, Singapore 117549 (e-mail: gemeng@nus.edu.sg; ephfm@nus.edu.sg). 


Haizhou Li, Department of Electrical and Computer and Engineering, National University of Singapore, Singapore 119077; Shenzhen Research Institute of Big Data, School of Data Science, The Chinese University of Hong Kong, Shenzhen 518172, China (email: haizhouli@cuhk.edu.cn).}
\thanks{This work is supported by Huawei Noah’s Ark Lab. We thank Liqun Deng (email: dengliqun.deng@huawei.com) for participation in discussions.
This work is also supported by the National Research Foundation Singapore under AI Singapore Programme (Award Number: AISG-GC-2019-001-2A and AISG2-TC-2022-004).  $^{*}$ Corresponding author.}
}

\maketitle

\begin{abstract}
Self-supervised pre-trained speech models were shown effective for various downstream speech processing tasks. Since they are mainly pre-trained to map input speech to pseudo-labels, the resulting representations are only effective for the type of pre-train data used, either clean or mixture speech. With the idea of selective auditory attention, we propose a novel pre-training solution called Selective-HuBERT, or SHuBERT, which learns the selective extraction of target speech representations from either clean or mixture speech. Specifically, SHuBERT is trained to predict pseudo labels of a target speaker, conditioned on an enrolled speech from the target speaker. By doing so, SHuBERT is expected to selectively attend to the target speaker in a complex acoustic environment, thus benefiting various downstream tasks. We further introduce a dual-path training strategy and use the cross-correlation constraint between the two branches to encourage the model to generate noise-invariant representation. Experiments on SUPERB benchmark and LibriMix dataset demonstrate the universality and noise-robustness of SHuBERT. Furthermore, we find that our high-quality representation can be easily integrated with conventional supervised learning methods to achieve significant performance, even under extremely low-resource labeled data.

\end{abstract}


\IEEEpeerreviewmaketitle

\vspace{-0.3cm}
\section{Introduction}

\IEEEPARstart{S}{elf-}supervised pre-training has gained much popularity among the researchers in recent years~\cite{mohamed2022self,ericsson2022self,qiu2020pre,jing2020self}. By pre-training the models on large amounts of unannotated speech data, not only is the cost of annotation saved, but the models also generate universal speech representations that are not specific to a particular task. 
Such universal representations can benefit a wide range of speech processing tasks, including speech recognition, speaker verification, acoustic word embeddings etc~\cite{schneider2019wav2vec,hsu2021hubert,chen2022large,lin2023self,fan2020exploring,yue2022self}, with finetuning of the pre-trained model~\cite{baevski2020wav2vec,baevski2022data2vec,sadhu2021wav2vec} or downstream task-specific layers~\cite{yang2021superb,tsai2022superb,huang2022investigating}.

\begin{figure}[t] 
\centering 
\includegraphics[width=\linewidth]{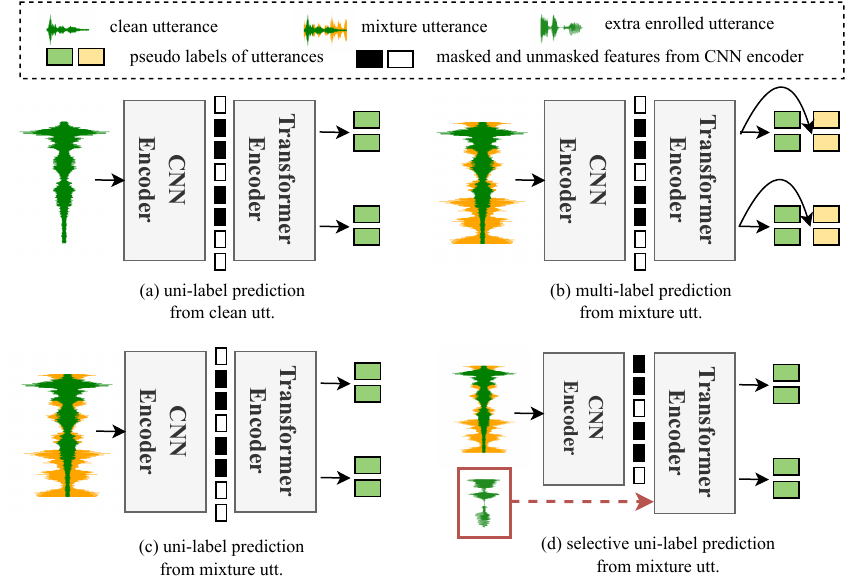} 
\caption{Illustration of four self-supervised pre-training frameworks on speech processing. (a) uni-label prediction from clean utterance. (b) Multi-label prediction from mixture utterance. (c) uni-label prediction from mixture utterance. (d) Our selective uni-label prediction from mixture utterance.
}
\label{fig:motivation}
\vspace{-1em}
\end{figure}

However, most advances have primarily been focused on tasks involving clean speech. When these pre-trained models are applied to mixture speech, their performance deteriorates substantially~\cite{ng2023hubert,wang2022wav2vec,zhu2023robust}. In real-world scenarios, it is unrealistic to assume that high-quality clean speech corpus is always readily available. 
To boost the noise robustness of pre-training models, recent researchers have explored 
pre-training with simulated mixture speech~\cite{chen2022wavlm,wang2023adapter,zhu2023robust,song2023exploring,hu2023wav2code}. In the speech community, the prevailing
self-supervised pre-training strategies can be mainly divided into two categories, including \textbf{mapping strategy} and \textbf{denoising strategy}. The former category, as denoted by (a) and (b) in Fig. \ref{fig:motivation}, can be conceptualized as a direct-mapping representation learning for all sources in the input speech. In other words, this strategy forces the model to predict single pseudo labels (i.e., uni-label) from input clean speech~\cite{hsu2021hubert}, or multiple pseudo labels (i.e., multi-label) corresponding to all sources in input mixture speech~\cite{fazel2023cocktail}. 
Nevertheless, models pre-trained with the mapping strategy generate representations that are biased towards the pre-training inputs, exhibiting a fundamental limitation in the adaptability for both clean and mixture speech downstream tasks. 
This prompts an imperative question: how to devise the pre-training process to obtain a highly robust speech representation suitable for both clean and mixture speech?

The denoising strategy, illustrated by Fig. \ref{fig:motivation}(c), provides a solution to this question. The denoising strategy integrates the denoising capability within the self-supervised pre-training process. This capability enables the model to naturally accept both clean and mixture as input for uni-label prediction. 
Although this strategy mitigates the inherent input bias in the mapping strategy,
it generally requires a prior definition of the primary speaker to avoid the speaker confusion problem. The primary speaker in this strategy is typically defined as individuals exhibiting either higher speech energy level~\cite{wang2023adapter} or longer speech duration~\cite{chen2022wavlm}, or as the sole speaker identity within noisy environments~\cite{ng2023hubert}. However, such definition naturally brings human assumptions, and may hurt the generated representations when the definition of the primary speaker in downstream tasks is inconsistent with the pre-training~\cite{huang2023self}.



In this work, we rethink the strengths and weaknesses of the current denoising strategy and propose SHuBERT, a new selective denoising strategy that improves upon it. As shown in Fig.~\ref{fig:motivation}(d), we adopt the same uni-label prediction as in Fig.~\ref{fig:motivation}(c) to preserve the representation’s universality on both clean and mixture speech. To eliminate the necessity for a primary speech definition, we use an enrolled speech from the target speaker, which is used to guide the model to identify and predict only the pseudo labels of the target speech, hence equipping the model with the selective attention capability.
This design is inspired by the selective attention mechanism observed in human perception~\cite{cherry1953some} and its relevance in the speech extraction tasks~\cite{ge2020spex+,ge2021multi,ge2022spex}. To further enhance the noise robustness, we propose a dual-path contrastive learning strategy, which builds upon the intuition that the speech content should be invariant to diverse background interferences.

\begin{figure*}[t] 
\centering 
\includegraphics[width=0.95\linewidth]{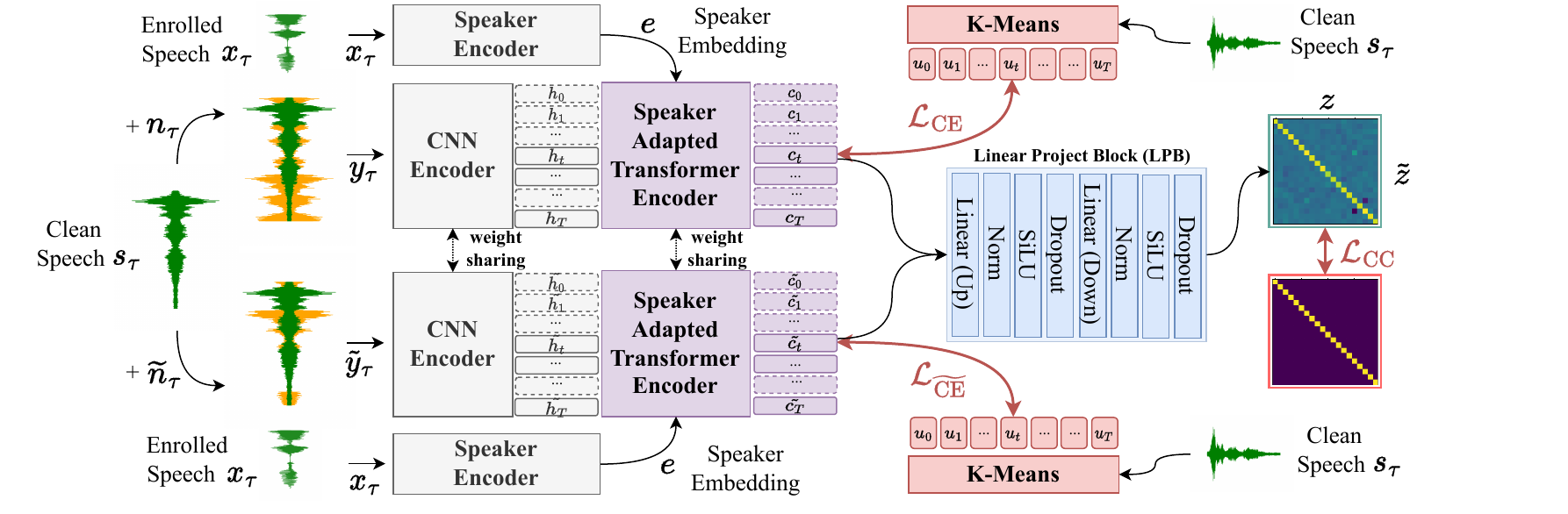} 
\caption{Illustration of our SHuBERT pipeline. A clean speech $s_{\tau}$ is first added with two different interferences to generate two mixture speech $y_{\tau}$ and $\widetilde{y}_{\tau}$, respectively. Given $y_{\tau}$, $\widetilde{y}_{\tau}$ and an extra enrolled speech $x_{\tau}$ from the target speaker, SHuBERT is pre-trained to generate representations $C$ and $\widetilde{C}$ that can predict the pseudo labels at the masked frames corresponding to $s_{\tau}$ and at the same time minimize the cross-correlation loss between $C$ and $\widetilde{C}$.
}
\label{fig:framework}
\vspace{-1em}
\end{figure*}

\vspace{-0.2cm}
\section{HuBERT}
Hidden Unit BERT (HuBERT) is a classic self-supervised speech pre-training framework. It benefits from an offline clustering (e.g., k-means) to generate target unit pseudo labels $U = [u_1, u_2, ..., u_T]$ from input clean speech $s_{\tau}$, where $T$ is the number of speech frames. Given the clean speech $s_{\tau}$, a convolutional neural network (CNN) $\mathcal{F}(\cdot)$ encodes $s_{\tau}$ into frame-level speech representation $H = [h_1, h_2, ..., h_T]$, which are then fed to a Transformer encoder $\mathcal{G}(\cdot)$ to get the contextual representation $C = [c_1, c_2, ..., c_T]$. During the pre-training process, the frame-level features $H$ should be masked randomly before they are passed to the Transformer encoder. The whole pipeline can be formalized as 
\begin{equation}
    C = [c_1, c_2, ..., c_T] = \mathcal{G}[\mathcal{M}(\mathcal{F}(s_{\tau}))],
\label{eq:hubert-pred}
\end{equation}
where $\mathcal{M}(\cdot)$ is the mask operation on frame-level representation. Finally, the model is trained to predict the pseudo labels of the masked frames using a cross-entropy (CE) loss function
\begin{equation}
    \mathcal{L}_{\text{CE}}(C, U) = \sum\nolimits_{t \in O} \log \, p_t (\, u_t \, | \, c_t),
\label{eq:hubert-ce-loss}
\end{equation}
where $O$ denotes the set of indices masked in $H$. Note that when applied to downstream tasks, $H$ is no longer masked and the obtained unmasked contextual representation $C$ is used.

\vspace{-0.2cm}
\section{SHuBERT Architecture}
Fig. \ref{fig:framework} illustrates the overall architecture of SHuBERT. SHuBERT is an extraction-based self-supervised learning framework, which aims to extract the target speaker's unique pseudo label from the mixture speech given a target speaker's enrolled speech. The main differences between HuBERT and SHuBERT lie in two aspects: Speaker-Adapted Transformer Encoder (SATE) and Dual-Path Contrastive Learning (DPCL). SATE aims to address \textit{how to utilize an enrolled speech to enhance the universality on both clean and mixture corpus}, while DPCL focuses on \textit{how to adopt dual-path pre-training to improve the noise robustness of the representations}.

\vspace{-0.4cm}
\subsection{Speaker Adapted Transformer Encoder}
\label{sec:SATE}

Speaker Adapted Transformer Encoder (SATE) is designed to generate the contextual representation $C = [c_1, c_2, ..., c_T]$ for the target speaker, conditioned on the speaker embedding $e$ from the enrolled speech and the masked frame-level representation $H^{m} = \mathcal{M}([h_1, h_2, ..., h_T])$ from mixture speech $y_{\tau} = s_{\tau} + n_{\tau}$, where $n_{\tau}$ is any background interferences (e.g., environmental noises or interfering speech from other speakers). This encoder consists of the Speaker Adapted Transformer Layer (SATL) and the Vanilla Transformer Layers (VTL). The difference between them is the layer normalization operation. VTL uses the traditional layer normalization to normalize the distributions of the intermediate representations $H^{m}$, while SATL adopts a conditional layer normalization operation that takes in an additional speaker embedding $e$ as input. The two normalization operations can be formulated as
\begin{gather}
    \hat{H}_{\text{VTL}} = \frac{H^{m} - E[H^{m}]}{\sqrt{Var[H^{m}] + \epsilon}} * \gamma + \beta, \\
    \hat{H}_{\text{SATL}} = \frac{H^{m} - E[H^{m}]}{\sqrt{Var[H^{m}] + \epsilon}} * [w(e) * \gamma + b(e)] + \beta,
\label{eq:layer_norm}
\end{gather}
where $E[H^{m}]$ and $Var[H^{m}]$ are the mean and variance of the input representation $H^{m}$, and $\gamma$ and $\beta$ are learnable weight and bias for applying the element-wise affine transformation. The $w$($\cdot$) and $b$($\cdot$) are linear projections that project the speaker embedding $e$ to the feature dimension of $H^{m}$.

Compared to HuBERT, i.e., Eq. (\ref{eq:hubert-pred}), the SATE incorporates a speaker embedding from the enrolled speech. This allows the model to exhibit the capacity for selective extraction of the target speaker's speech representation from diverse input, e.g., clean or mixture speech. Let $\mathcal{G}_{\text{SATE}}(\cdot)$ denote the SATE, the selective extraction pipeline is formulated as
\begin{equation}
    C = [c_1, c_2, ..., c_T] = \mathcal{G}_{\text{SATE}}[\mathcal{M}(\mathcal{F}(y_{\tau})), \, e].
\label{eq:single-path-pred}
\vspace{-0.1cm}
\end{equation}

\vspace{-0.4cm}
\subsection{Dual-Path Contrastive Learning}
Our motivation is that a noise-robust model should generate similar representations for a speech signal $s_{\tau}$ with the presence of arbitrary background interferences (including both noises and interfering speech), and the generated representations should primarily reflect the content of $s_{\tau}$. This motivates us to adopt dual-path contrastive training to encourage noise-invariant contextual representation generation.

As shown in Fig. \ref{fig:framework}, we introduce two background interferences $n_{\tau}$ and $\widetilde{n}_{\tau}$ into clean speech $s_{\tau}$, and design two branches to take in the two corrupted versions $y_{\tau}=s_{\tau}+n_{\tau}$ and $\widetilde{y}_{\tau}=s_{\tau}+\widetilde{n}_{\tau}$. Similar to the procedure for extracting $C$ from $y_{\tau}$ in Eq. (\ref{eq:single-path-pred}), $\widetilde{C}$ is also extracted through the weight-shared CNN $\mathcal{F}(\cdot)$ and SATE $\mathcal{G}_{\text{SATE}}(\cdot)$. This is denoted as
\begin{equation}
    \widetilde{C} = [\widetilde{c}_1, \widetilde{c}_2, ..., \widetilde{c}_T] = \mathcal{G}_{\text{SATE}}[\mathcal{M}(\mathcal{F}(\widetilde{y}_{\tau})), \, e].
\label{eq:dual-path-pred}
\end{equation}

Next, the contextual representations $C$ and $\widetilde{C}$ are passed through a shared Linear Projection Block (LPB)~\cite{ng2023hubert} and randomly downsampled to $Z = [z_{1}, z_{2}, ..., z_{N}]$ and $\widetilde{Z} = [\widetilde{z}_{1}, \widetilde{z}_{2}, ..., \widetilde{z}_{N}]$ respectively, where $N$ is the number of frames sampled. Here, the frame-level downsampling is to mitigate the potential overestimation of variability introduced by frame-level correlation. We then adopt the Barlow Twins' cross-correlation (CC) loss in~\cite{zbontar2021barlow} as the feature-level contrastive loss between $Z$ and $\widetilde{Z}$. Specifically, we push the feature-level cross-correlation (CC) matrix towards the identity matrix. This means maximising the speech content agreement (pushing the diagonal elements towards 1), and minimising other background interferences (pushing the non-diagonal elements towards 0). Hence, this loss is formulated as 
\vspace{-0.1cm}
\begin{gather}
    {R_{ij}}^2 = \frac{\sum_{n} z_{n,i} \widetilde{z}_{n,j}}{\sqrt{\sum_{n} (z_{n,i})^2} \sqrt{\sum_{n} (\widetilde{z}_{n,i})^2}}, \\
    \mathcal{L}_{\text{CC}} (Z, \widetilde{Z}) = \sum_{i} (1-R_{ii})^2 + \lambda \sum_{i} \sum_{j \neq i} {R_{ij}}^2
\label{eq:cc_loss}
\end{gather}
where $n$ denotes the index of frames sampled, $i$ and $j$ refer to the dimensional position of the frame-level representations. $R$ denotes the feature-level cross-correlation matrix between $Z$ and $\widetilde{Z}$, and $\lambda$ denotes a balance parameter.

Finally, we incorporate the contrastive learning process into our extraction-based pipeline and form our overall loss as
\begin{equation}
    \mathcal{L}_\text{DPCL} = \mathcal{L}_{\text{CE}}(C, U) + \mathcal{L}_{\widetilde{\text{CE}}}(\widetilde{C}, U) + \mathcal{L}_{\text{CC}}(Z, \widetilde{Z})
\label{eq:overall_loss}
\vspace{-0.1cm}
\end{equation}


\section{Experiments}
\subsection{Datasets}
Following~\cite{wang2023adapter}, we conduct the pre-training experiments on the data simulated from the 960-hour LibriSpeech~\cite{panayotov2015librispeech} and the DNS Challenge's noise datasets~\cite{reddy2020interspeech}. In practice, the simulated mixture speech is generated by a dynamic mixing strategy~\cite{zhang2023weakly}, including single-talker noisy speech, two-talker overlapped speech, and noisy two-talker overlapped speech. The required enrolled speech of the target speaker is a randomly selected utterance that is different from the utterance in the mixture. The speaker embedding for the Speaker Adapted Transformer Encoder (SATE) can be generated from the corresponding enrolled speech with any pre-trained speaker models~\cite{9747021,desplanques2020ecapa,snyder2018x}. Here we use the pre-trained CAM++~\cite{zheng20233d}. All audios are sampled at 16kHz.


\vspace{-0.3cm}
\subsection{Experimental Setup}
\label{sec:experimental-setup}
All pre-training experiments are conducted using the Fairseq toolkit~\cite{ott2019fairseq}. Our SHuBERT's parameters are initialized from the pre-trained HuBERT Base,
with the exception of $w$ and $b$ of Speaker Adapted Transformer Layer (SATL). SATL can replace any Vanilla Transformer Layer (VTL). Here, we only study the replacement of the 1st VTL. We pre-train SHuBERT for 400k steps with a learning rate of 7e-5. The pseudo labels for pre-training are generated by the 500-centroid K-means model trained on representations from the 9th transformer layer in the HuBERT Base model~\cite{fazel2023cocktail}. Other hyperparameters for the model are kept the same as in~\cite{hsu2021hubert}. The hyperparameters for Dual-Path Contrastive Learning (DPCL) follow~\cite {ng2023hubert}.


Since our model is based on HuBERT architecture, we select a series of HuBERT models as baselines: 
1) \textbf{HuBERT Base} (Fig.~\ref{fig:motivation}(a)): Pre-trained on only clean data simulated from 960-hour Librispeech; 
2) \textbf{HuBERT PSD Base} (Fig.~\ref{fig:motivation}(c)): Adopts the primary-speaker denoising scheme~\cite{chen2022wavlm}, where the primary speaker is defined as the speaker with a longer speech duration and as the sole speaker in noisy environments. The data is also simulated from 960-hour LibriSpeech and DNS noises. It has the same structure as HuBERT; 
3) \textbf{C-HuBERT}~\cite{fazel2023cocktail} (Fig.~\ref{fig:motivation}(b)): Pre-trained on multi-talker mixtures (more than 2 speakers) simulated from 960-hour Librispeech. It modifies the HuBERT architecture to include multiple prediction heads to predict pseudo labels of each source in the input mixtures.




\vspace{-0.4cm}
\subsection{Universal Speech Representation Evaluation on SUPERB}
To verify the universality of the learned speech representation by SHuBERT on both clean and mixture speech, we conduct experiments on the Speech processing Universal PERformance Benchmark (SUPERB)~\cite{tsai2022superb}. We choose phoneme recognition (PR) and automatic speech recognition (ASR) as clean speech tasks, and speech enhancement (SE) and separation (SS) as the mixture speech tasks. The evaluation metrics for PR and ASR are phoneme error rate (PER) and word error rate (WER). Perceptual evaluation of speech quality (PESQ) and short-time objective intelligibility (STOI) are reported for SE, and scale-invariant signal-to-distortion ratio improvements (SI-SDRi) is reported for SS. Following the basic SUPERB settings, the pre-trained models are frozen and only the downstream layers are fine-tuned. Note that the speaker embedding in clean speech tasks is directly extracted from the input speech, while in mixture speech tasks, it is extracted from an enrolled speech. On top of HuBERT baselines mentioned in Section~\ref{sec:experimental-setup}, we also compare SHuBERT without and with DPCL, denoted by 1-path and 2-path respectively.


Table~\ref{tbl:superb} reports the evaluation results for SHuBERT and HuBERT-related baselines on SUPERB. With the experimental observations from Table~\ref{tbl:superb}, we draw the following conclusions: 1) the mapping strategy produces representations that are inevitably biased towards the pre-training inputs and hence, suffer from limited adaptability to allow simultaneous application on both clean and mixture speech tasks. This can be seen from the results of HuBERT and C-HuBERT, in which they only perform well in either clean (HuBERT) or mixture (C-HuBERT) tasks but not both. 2) The denoising strategy provides more universal representations that can be applied for both clean and mixture speech tasks. From the results, we can see that both SHuBERT variants obtain comparable or superior performances across clean and mixture tasks.  3) Enhanced noise robustness in the pre-trained SHuBERT model is achieved through DPCL, as evidenced by the better performances of SHuBERT (2-path) over SHuBERT (1-path) across all the tasks.


\begin{table}[t]
\centering
    \scriptsize
    \def\arraystretch{1.01}
    \setlength{\tabcolsep}{1.55pt}
    \setlength{\abovetopsep}{0pt}
    \setlength\belowbottomsep{0pt} 
    \setlength\aboverulesep{0pt} 
    \setlength\belowrulesep{0pt}
\caption{Universal Speech Representation Evaluation on SUPERB. We evaluate on two clean speech tasks and two mixture tasks.}
\vspace{-0.5em}
\scalebox{1}{
\begin{tabular}{l|c|c|cc|ccc}
\toprule
\multirow{3}{*}{\textbf{Methods}} 
& \multirow{3}{*}{\textbf{\#Param}}
& \multirow{3}{*}{\textbf{Strategy}}
& \multicolumn{2}{c|}{\textbf{Clean Tasks}} 
& \multicolumn{3}{c}{\textbf{Mixture Tasks}}  \\
\cmidrule{4-8}
&  &  & PR & ASR  & \multicolumn{2}{c}{SE} & SS \\ \cmidrule{4-8}

& &  & PER$\downarrow$ & WER$\downarrow$  & PESQ$\uparrow$ & STOI$\uparrow$ &SI-SDRi$\uparrow$ \\
\midrule
FBANK & -- & -- & 82.01 & 23.18 & 2.55 & 93.60 & 9.23 \\
\midrule
HuBERT Base~\cite{hsu2021hubert} & 94.7M & Fig. \ref{fig:motivation}(a) & 5.41  & 6.42  & 2.58 & 93.90 & 9.36  \\

C-HuBERT Base~\cite{fazel2023cocktail} & 96.0M & Fig. \ref{fig:motivation}(b) & 6.14  & 7.38 & \textbf{2.63} & 94.00 & \textbf{11.08}  \\

HuBERT PSD Base & 94.7M & Fig. \ref{fig:motivation}(c) & 5.66  & 6.35 & 2.58 & 93.83 & 9.39  \\
\midrule


SHuBERT (1-path) & 95.3M & Fig. \ref{fig:motivation}(d) & 5.50  & 6.40 & 2.62 & 94.11 & 10.38  \\


SHuBERT (2-path) & 100.0M & Fig. \ref{fig:motivation}(d) & \textbf{5.04}  & \textbf{6.28} & 2.62 & \textbf{94.19} & 10.53  \\

\toprule

\end{tabular}}
\label{tbl:superb}
\vspace{-1.5em}
\end{table}

\vspace{-0.4cm}
\subsection{Integration with Conventional Supervised Model}
To evaluate the compatibility of our SHuBERT, we consider integrating pre-trained models with conventional supervised models (e.g., ConvTasNet) and evaluate on LibriMix~\cite{cosentino2020librimix}. We believe that the pre-trained representation can provide richer contextual information to help current supervised learning models. Thus, we concatenate the speech representation from pre-training models to embedded features from the ConvTasNet encoder in practice.

Table~\ref{tbl:low-resource} reports the results for the low- and high-resource setup, where pre-trained models are fine-tuned on 1\%, 10\% and 100\% of labeled training data. From Table~\ref{tbl:low-resource}, we first observe that the supervised model (i.e., ConvTasNet) benefits from the pre-trained contextual representations, especially under low-resource setting (e.g., 1\%). This can be explained as pre-trained contextual representations can provide more noise-robust information, such as content-related cues. In addition, we find that ConvTasNet with our representation achieves the best performances on both single-talker noisy speech (i.e., mixture-single) and two-talker overlapped speech (i.e., mixture-clean). With pre-training on mixture speech, HuBERT PSD and SHuBERT are significantly better than HuBERT. 
Besides, compared with HuBERT PSD, our SHuBERT representation achieves similar results in the mixture-single case, and achieves 0.4dB+ SI-SDRi improvement in the mixture-clean case. The reason is that HuBERT PSD faces definition mismatch issues in the mixture-clean case, and SHuBERT can naturally avoid this issue. The above discussions further verify the effectiveness of our SHuBERT.

\begin{table}[t]
\centering
    \scriptsize
    \def\arraystretch{1.01}
    \setlength{\tabcolsep}{1.55pt}
    \setlength{\abovetopsep}{0pt}
    \setlength\belowbottomsep{0pt} 
    \setlength\aboverulesep{0pt} 
    \setlength\belowrulesep{0pt}
\caption{Results on low resource setups, including 1\%, 10\%, and 100\% of training data. The downstream separation model is ConvTasNet. 100\% of traning data contains 13,900 utterances.}
\vspace{-0.5em}
\scalebox{1}{
\begin{tabular}{l|c|c|cc|cc}
\toprule
\textbf{Train}
& \textbf{Pre-training} & \textbf{Downstream}
& \multicolumn{2}{c|}{\textbf{Mixture-Single}} 
& \multicolumn{2}{c}{\textbf{Mixture-Clean}}  \\
\cmidrule{4-7}
\textbf{Data}&\textbf{Models} &\textbf{Network} & SDRi$\uparrow$ & SI-SDRi$\uparrow$  & SDRi$\uparrow$ & SI-SDRi$\uparrow$ \\
\midrule
\multirow{4}{*}{1\%}&  --  &\multirow{4}{*}{ConvTasNet} & 7.07  & 6.52  & 3.05 & 2.59  \\

& HuBERT Base~\cite{hsu2021hubert}   & & 7.83  & 7.22 & 4.45 & 3.97  \\

& HuBERT PSD Base  & & \textbf{8.28} & 7.73 & 7.72  & 7.26   \\

& SHuBERT & & 8.27 & \textbf{7.74} & \textbf{8.11}  & \textbf{7.68}  \\
\midrule

\multirow{4}{*}{10\%}&  --  &\multirow{4}{*}{ConvTasNet} & 9.94  & 9.52  & 9.73 & 9.16  \\

& HuBERT Base~\cite{hsu2021hubert}  & & 9.93 & 9.37 & 11.08 & 10.67  \\

& HuBERT PSD Base  & & 10.34 & 9.80 & 12.59 & 12.26  \\

& SHuBERT & & \textbf{10.38}  & \textbf{9.89} & \textbf{13.51} & \textbf{13.20} \\
\midrule

\multirow{4}{*}{100\%}&  --  &\multirow{4}{*}{ConvTasNet} & 11.35  & 10.77  & 14.53 & 14.12  \\

& HuBERT Base~\cite{hsu2021hubert}  &  & 11.46  & 10.83 & 14.99 & 14.62  \\

& HuBERT PSD Base  & & 11.66  & 11.07 & 16.14 & 15.79  \\

& SHuBERT &  & \textbf{11.68}  & \textbf{11.09} & \textbf{16.50} & \textbf{16.18} \\

\toprule

\end{tabular}}
\label{tbl:low-resource}
\vspace{-1em}
\end{table}

\begin{table}[t]
\centering
    \scriptsize
    \def\arraystretch{1.2}
    \setlength{\tabcolsep}{1.55pt}
    \setlength{\abovetopsep}{0pt}
    \setlength\belowbottomsep{0pt} 
    \setlength\aboverulesep{0pt} 
    \setlength\belowrulesep{0pt}
\caption{Results on Multi-talker Speech Recognition.}
\vspace{-0.5em}
\scalebox{1}{
\begin{tabular}{l|c|c}
\toprule
\textbf{Pre-training Methods} &  \textbf{Spk. Embedding} & \textbf{WER (\%)\,} \\

\midrule

\multirow{2}{*}{HuBERT Base~\cite{hsu2021hubert}} & \xmark & 13.04  \\
 & \cmark & 13.58  \\
\hline 

\multirow{2}{*}{HuBERT PSD Base} & \xmark &  9.65 \\
 & \cmark & 9.34 \\
\hline 

SHuBERT & \cmark & \textbf{8.26} \\

\toprule

\end{tabular}}
\label{tbl:multi-talker-asr}
\vspace{-1.5em}
\end{table}

\vspace{-0.4cm}
\subsection{Evaluation on Multi-talker Speech Recognition}
To further verify our representation's noise-robustness, we conduct experiments on highly-overlapped multi-talker speech recognition tasks by following the utterance-based evaluation settings in \cite{huang2023adapting}. Table~\ref{tbl:multi-talker-asr} shows the WER for HuBERT, HuBERT PSD and SHuBERT. For both HuBERT and HuBERT PSD, we also consider the cases where the speaker embeddings are injected into the Transformer encoder in the same way as SHuBERT. As shown in Table~\ref{tbl:multi-talker-asr}, HuBERT exhibits suboptimal performance in highly overlapped mixture data, resulting in a WER of 13.04\%, and the performance is not improved even with the injection of speaker embeddings. 
In contrast, HuBERT PSD and SHuBERT outperform HuBERT by a large margin. HuBERT PSD achieves a WER of 9.65 which is a notable 26\% reduction, even in the absence of the speaker embeddings. With the incorporation of speaker embeddings, the WER is further reduced to 9.34. Notably, SHuBERT still outperforms HuBERT PSD, achieving a WER of 8.26, which is a 36.6\% reduction from HuBERT Base. 





\vspace{-0.2cm}
\section{Conclusion}
This paper presents SHuBERT, a pre-training strategy that adapts the pre-training models to noisy and mixture speech environments with a strong selective module and enhanced noise resilience. We proposed two major modifications to the pre-training models: the Speaker Adapted Transformer Encoder (SATE) and the Dual-Path Contrastive Learning (DPCL) strategy. A series of experiments have been carried out to show the superiority in handling both clean and mixture speech.

\bibliographystyle{IEEEtran}
\bibliography{pretraining}

\end{document}